\documentclass[aps,pra,twocolumn,floats,amsmath,amssymb,superscriptaddress]{revtex4-1}
\usepackage[utf8]{inputenc}
\usepackage{graphicx}
\usepackage{epsfig}
\usepackage{amsfonts}
\usepackage{amsmath}
\usepackage{natbib}
\usepackage{multirow}
\usepackage{bm}
\usepackage{epstopdf}
\usepackage{ulem}
\usepackage{color}
\definecolor{/green}{rgb}{0.0, 0.72, 0.92}

\newcommand{\ie}{{\it i.e.}, }
\newcommand{\eg}{{\it e.g.}, }

\begin{document}
\title{Two-dimensional electron gas at LaInO$_3$/BaSnO$_3$ interfaces controlled by a ferroelectric layer}
%%%%%%%%%%%%%%%%%%%%%%%%%%%%%%%%%%%%%%%%%%%%%%%%%%%%%%%%%%%%%
\author{Le Fang}
\affiliation{Materials Genome Institute, International Center for Quantum and Molecular Structures, Physics Department, Shanghai University, 200444 Shanghai, China}
\affiliation{Institut f\"{u}r Physik and IRIS Adlershof, Humboldt-Universit\"{a}t zu Berlin, 12489 Berlin, Germany}
\author{Wahib Aggoune}
\affiliation{Institut f\"{u}r Physik and IRIS Adlershof, Humboldt-Universit\"{a}t zu Berlin, 12489 Berlin, Germany}
\author{Wei Ren}
\email{renwei@shu.edu.cn}
\affiliation{Materials Genome Institute, International Center for Quantum and Molecular Structures, Physics Department, Shanghai University, 200444 Shanghai, China}
\affiliation{Shanghai Key Laboratory of High Temperature Superconductors and State Key Laboratory of Advanced Special Steel, Shanghai University, Shanghai 200444, China}
\author{Claudia Draxl}
\email{claudia.draxl@physik.hu-berlin.de} 
\affiliation{Institut f\"{u}r Physik and IRIS Adlershof, Humboldt-Universit\"{a}t zu Berlin, 12489 Berlin, Germany}
\affiliation{European Theoretical Spectroscopic Facility (ETSF)}

\begin{abstract}%
With the example of LaInO$_{3}$/BaSnO$_3$, we demonstrate how both density and distribution of a two-dimensional electron gas (2DEG) formed at the interface between these perovskite oxides, can be efficiently controlled by a ferroelectric functional material. A polarization induced in a BaTiO$_3$ layer pointing toward the interface enhances the polar discontinuity which, in turn, significantly increases the 2DEG density and confinement, while, the opposite polarization depletes the 2DEG. Our predictions and analysis, based on first-principles calculations, can serve as a guide for designing such material combinations to be used in electronic devices.
\end{abstract}
 
\date{\today}
\maketitle
A two-dimensional electron gas (2DEG) at the interface between two oxide perovskites was first observed at the combination of the polar material LaAlO$_3$ and the nonpolar compound SrTiO$_3$~\cite{Ohtomo+04n}. In such a combination, electronic reconstruction occurs to compensate the polar discontinuity at the (LaO)$^+$/(TiO$_2$)$^0$ interface, forming a 2DEG within the SrTiO$_3$ side~\cite{Nakagawa+06nm}. The latter can reach a free charge-carrier density of $\sim$3.3$\times$10$^{14}$ cm$^{-2}$ (0.5 electrons per unit-cell area)~\cite{Huijben+09am} and can be confined within a few nanometers, offering the possibility to fabricate nanoscale electronic circuits in a narrow conducting path ($\sim$2 nm)~\cite{cen+2009Sc,cen+2008NM}. As such, 2DEGs at oxide interfaces have attracted tremendous interest due to their enormous potential in the next generation of electronic devices. 
 
Practical applications of such interfaces, \eg in field-effect transistors (FET), require to control the 2DEG. To tune the 2DEG properties in terms of density and confinement, several external stimuli have been proposed, such as light illumination~\cite{irvin+2010NP}, strain~\cite{bark+2011PNAS}, or an electric ﬁeld~\cite{Thiel+06sc,cen+2008NM}. Also, addition of a functional ferroelectric layer has been suggested that allows for switching its polarization ~\cite{tra+2013AM,kim+2013AM}. Such a method can be quite effective as the microstructure of the interface itself is preserved. Indeed, a tunable 2DEG has been observed experimentally in a heterostructure consisting of LaAlO$_3$/SrTiO$_3$ and a ferroelectric Pb(Zr,Ti)O$_3$ layer~\cite{tra+2013AM}. Depending on the polarization direction, the 2DEG could be reversibly turned \textit{on} and \textit{off} with a large \textit{on/off} ratio (\textgreater 1000) in a non-volatile manner. X-ray photoelectron spectroscopy and cross-sectional scanning tunneling microscopy provided evidence for the modulation being caused by changes of the interfacial electronic structure through the ferroelectric polarization~\cite{tra+2013AM}. 

In recent years, perovskite oxide interfaces have developed rapidly, but the realization of functionality at room temperature is still challenging. The free charge-carrier mobility is a key parameter influencing the performance of any electronic device. At the LaAlO$_3$/SrTiO$_3$ interface, it can reach 10$^4$ cm$^{2}$(Vs)$^{-1}$ at low temperature, but it decreases to 1 cm$^{2}$(Vs)$^{-1}$ at ambient conditions~\cite{Ohtomo+04n,can+2016Nc}. The intrinsic critical reason for this lies in the partially occupied Ti-3$d$ states that exhibit low dispersion, \textit{i.e.}, large effective electronic masses. Furthermore, scattering of electrons within these bands reduce the mobilities through electro-acoustic coupling. As a consequence, SrTiO$_{3}$-based heterostructures have significant limitations in practical applications in electronic devices.

Cubic BaSnO$_{3}$ is attracting great interest~\cite{kim+2015APL,Aggoune+BSO} due to its extraordinary high mobilities achieved at room temperature, reaching about 320 cm$^{2}$(Vs)$^{-1}$~\cite{Hkim+12ape,paudel+17prb}. This value is two orders of magnitude larger than that of SrTiO$_{3}$~\cite{koba+2015ACS}, and even the highest ever measured in transparent conducting oxides (TCOs). The experimental realization of coherent interfaces with LaInO$_3$, exhibiting proper lattice matching~\cite{Martina+20prm,Aggoune+LIO}, makes the LaInO$_3$/BaSnO$_3$ (LIO/BSO) system a most promising combination to overcome the limitations of LaAlO$_3$/SrTiO$_3$~\cite{kim+2015APL,agg+21npj,Martina+21}. Therefore, it is important to explore how to control its 2DEG by an external stimulus like a ferroelectric functional layer.

%%%%%%%%%%%%%%%%%%%%%%%%
\begin{figure*}[hbt]
 \begin{center}
\includegraphics[width=.96\textwidth]{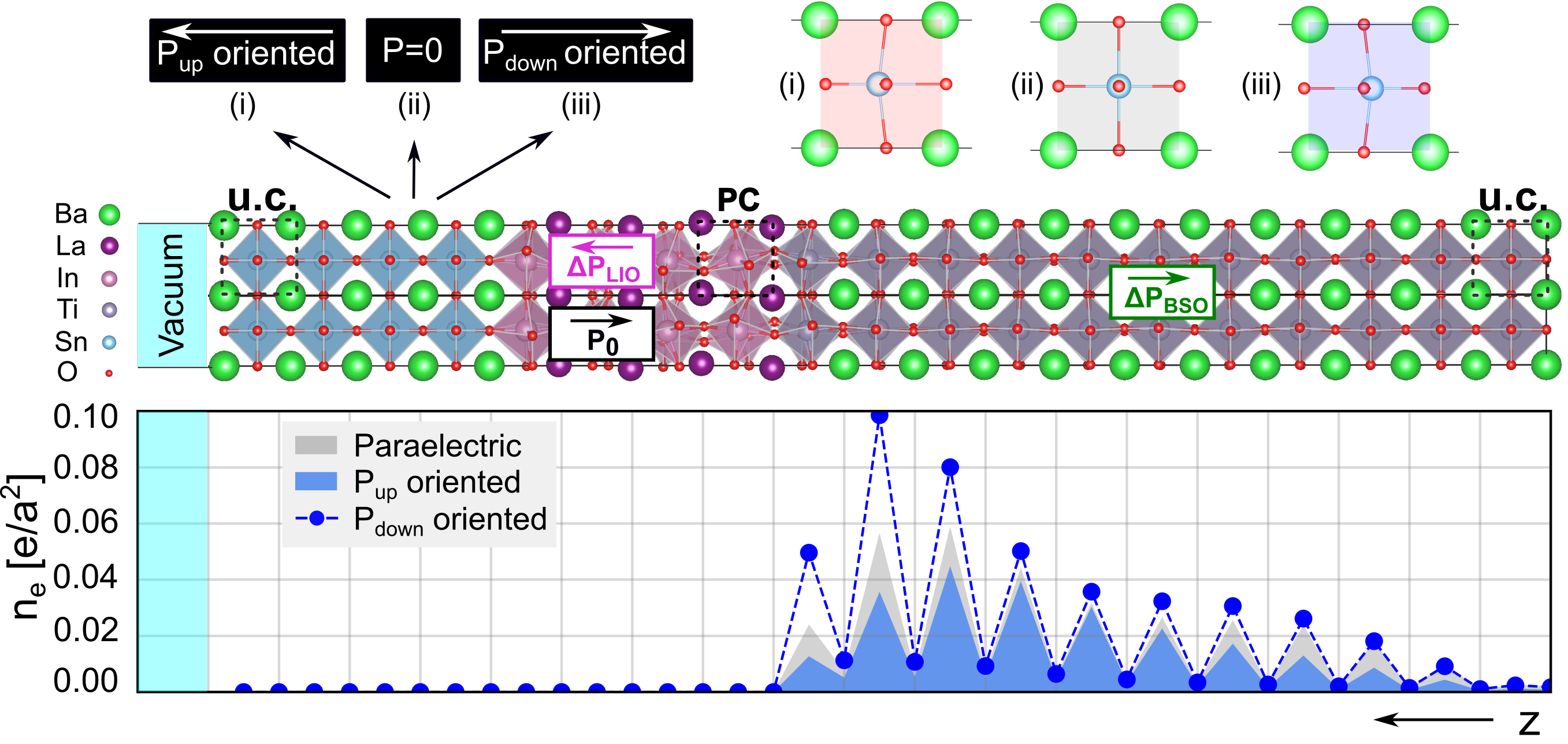}%
\caption{Distribution of the 2D electron charge density in the BaSnO$_3$ side along the $z$ direction for polarizations in BaTiO$_{3}$ oriented (i) outward (P$_{\textrm{up}}$) and (iii) toward (P$_{\textrm{down}}$) the interface, as well as the (ii) paraelectric case (P=0). The structural model of the BTO/LIO/BSO heterostructure is shown above. P$_{0}$ is the formal polarization oriented from the (InO$_{2}$)$^{-1}$ plane towards the (LaO)$^{+1}$ plane at the interface. Polarizations due to structural distortions within LaInO$_{3}$ ($\Delta P_{\mathrm{LIO}}$, magenta) and BaSnO$_{3}$ ($\Delta P_{\mathrm{BSO}}$, green) are indicated. The BaSnO$_{3}$ and BaTiO$_{3}$ unit cells (u.c) and the pseudocubic LaInO$_{3}$ unit cell (PC) are marked.}
\label{Fig-1}
 \end{center}
\end{figure*}
%%%%%%%%%%%%%%%%%%%%%%%%%%%%%%%%%%%%%%%%%%%%%%%%%

To do so, state-of-the-art theory can play an important role by getting insight into the mechanisms driving the formation of 2DEGs. In this Letter, we predict by first-principles calculations that the 2DEG formed at the LIO/BSO interface can be tuned by a ferroelectric material and we rationalize the miscroscopic mechanism behind it. We choose the tetragonal perovskite BaTiO$_3$, as it exhibits a polarization oriented along the [001] direction at room temperature, providing the possibility for pointing either toward or outward the interface~\cite{fred+2015PRB}. Moreover, as this ferroelectric material is made of neutral planes \textit{i.e.} is nonpolar in nature, it is ideally suited for studying purely the role of ferroelectricity~\cite{fred+2015PRB} in tuning the 2DEG at the LIO/BSO interface. We demonstrate that switching the polarization direction allows for accumulating or depleting charge at the interface. Analyzing the so caused changes in the polar discontinuity as well as in the valence and conduction band edges, we obtain insight into the microscopic mechanism behind the formation of a controllable 2DEG at such polar/nonpolar interface. Overall, we find a gradual variation of the 2DEG density by varying the amount of polarization. We also show how the material reacts to the ferroelectric layer with structural relaxations. These predictions open possibilities for designing oxide heterostructures with tailored characteristics.

In Fig.~\ref{Fig-1}, we depict the BaTiO$_3$/LaInO$_3$/BaSnO$_3$ (BTO/LIO/BSO) heterostructure, investigated in this work. The LIO/BSO interface is composed of four pseudocubic LaInO$_3$ and 11 BaSnO$_3$ unit cells, which is enough to capture the extension of the structural deformations as well as the 2DEG distribution away from the interface. The structural and electronic properties are calculated using density-functional theory (DFT) within the generalized gradient approximation (GGA) in the PBEsol parameterization~\cite{PBEsol+08prl} for exchange-correlation effects. These calculations are performed using FHI-aims~\cite{FHI-aims}, an all-electron full-potential package, employing numerical atom-centered orbitals. The first two BaSnO$_3$ unit cells are fixed to the bulk structure to simulate the bulk-like interior of the substrate. A vacuum layer ($\sim$140 \AA) along with a dipole correction are considered in the out-of plane direction [001] in order to prevent unphysical interactions between neighboring replica. For computing the electronic properties, a 20 $\times$ 20 $\times$ 1 $\textbf{k}$-grid is adopted for all systems. More details related to convergence and structural relaxation for the pristine materials and their interfaces can be found in Refs.~\onlinecite{Fang+long,agg+21npj}.

The interfacial polar discontinuity leads to the formation of a 2DEG at the BaSnO$_{3}$ side in order to compensate it. Upon increasing the LaInO$_3$ thickness (up to eight pseudocubic unit cells), its density increases, reaching a value of about 0.5 electrons per $a^{2}$ ($a$ being the BaSnO$_{3}$ lattice parameter~\cite{agg+21npj}). The geometry considered in this work (Fig.~\ref{Fig-1}) exhibits a 2DEG density of about 0.13 e/$a^{2}$~\cite{Fang+long,agg+21npj}. We note that this much smaller value is not only attributed to the effect of the thickness but also to the structural distortions within the LaInO$_3$ block. The latter reduce the polar discontinuity, hampering the charge transfer and the formation of a high-density 2DEG~\cite{Fang+long,agg+21npj}. Such a low density is ideal for exploring how the ferroelectric overlayer can tune and enhance it.
%%%%%%%%%%%%%%%%%%%%%%%%%%%%%%%%%%%%%%%%%%%%%%%%%%%
\begin{figure*}[ht]
	\begin{center}
		\includegraphics[width=.9\textwidth]{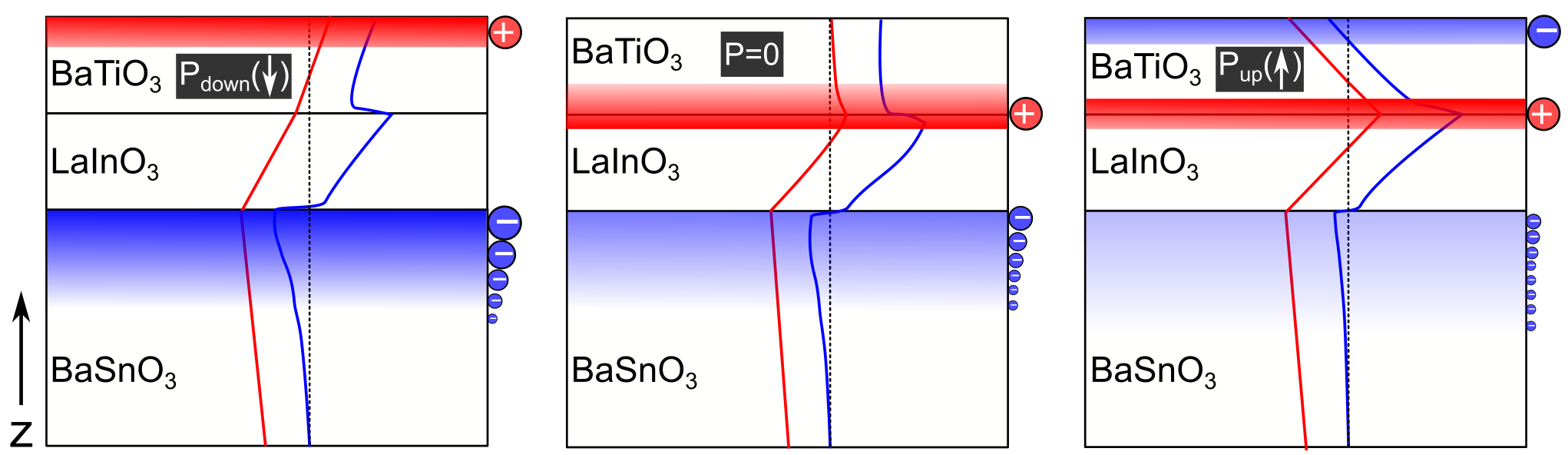}%
		\caption{Schematic diagram of the distribution of the electron and hole charges within the BTO/LIO/BSO heterostructures for P$_{\textrm{down}}$ (left), P=0 (middle), and P$_{\textrm{up}}$ polarization (right). The changes in the edges of the valence band (VBM) and conduction band (CBM), obtained from the analysis of the local density of states~\cite{Fang+long}, are also highlighted. The dashed black line indicates the Fermi level.}
		\label{Fig-2}
	\end{center}
\end{figure*}
%%%%%%%%%%%%%%%%%%%%%%%%%%%%%%%%%%%%%%%  

The in-plane lattice constants of the BTO/LIO/BSO heterostructures are fixed to $\sqrt{2}a_{\mathrm{BSO}}$, where a$_{BSO}$=4.119~\AA\ is obtained for cubic BaSnO$_3$ \cite{Fang+long}. This implies a constraint for the BaTiO$_{3}$ block whose in-plane spacing is a$_{BTO}$=3.954~\AA~\cite{Fang+long}. For the BaTiO$_{3}$ bulk structure, this leads to a tetragonal distortion with a $c/a$ ratio of 0.940 and a ferroelectric polarization of about 0.10 C/m$^2$ along the [001] direction~\cite{Fang+long}, compared to $c/a=$\,1.011 and a polarization of 0.27 C/m$^2$ in the pristine material~\cite{li+1998Asc,zhong+1994PRL}.

The magnitude of the ferroelectric polarization reported here is estimated using the cation-anion displacements and the Born effective charges (Z$^{*}$) computed for the bulk material. The same method is used for the BaTiO$_{3}$ block within the heterostructures. Z$^{*}$ are computed with the Berry-phase approach~\cite{berryPhase+93prbr} using \texttt{exciting}~\cite{gula+14jpcm}, an all-electron full-potential code, implementing the family of (L)APW+LO (linearized augmented planewave plus local orbital) methods. More details will be provided in a forthcoming publication~\cite{Fang+long}.

Including a BaTiO$_3$ overlayer of four unit cells thickness, we select the (BaO)$^0$ termination [Fig.~\ref{Fig-1} (center)] to avoid mid-gap states that would arise from TiO$_{2}$ termination~\cite{Padilla+97prb}. Three different scenarios are considered in the BaTiO$_3$ block. These are (i) polarization outward of the LIO/BSO interface (termed P$_{\textrm{up}}$), (ii) the paraelectric case (P=0), and (iii) polarization toward the interface (termed P$_{\textrm{down}}$) [Fig.~\ref{Fig-1} (top)]. The polarization magnitude is controlled by tuning the cation-anion displacements within the BaTiO$_3$ block. With this approach, we mimic the experimental situations where the ferroelectric polarization (orientation and amount) is controlled by applying an electric voltage~\cite{kim+2013AM}. For all considered polarizations, we fix the BaTiO$_3$ overlayer and optimize the BaSnO$_3$ and LaInO$_3$ blocks of the BTO/LIO/BSO heterostructures. In this way, we can analyze how the 2DEG density and its distribution as well as the atomic structure at the interface react to the ferroelectric polarization and its direction.

 Figure~\ref{Fig-1} depicts the distribution of the 2DEG accumulated within the BaSnO$_{3}$ block that results from the electronic reconstruction to compensate the polar discontinuity at the LIO/BSO interface~\cite{agg+21npj}. For case (ii), the centrosymmetric BaTiO$_3$ structure, a similar 2DEG distribution is found as in the pure LIO/BSO interface, however with an increased density of up to about 0.37 e/a$^2$ compared to 0.13 e/a$^2$ in LIO/BSO. This is mainly related to the additional charge transfer from BaTiO$_3$ to BaSnO$_3$ as will be described in more detail elsewhere~\cite{Fang+long}. Analyzing the structure, we find that the presence of BaTiO$_{3}$ reduces the structural distortions within LaInO$_{3}$ (specifically in the InO$_{2}$ layer closest to BaTiO$_{3}$). This enhances the polar discontinuity at the LIO/BSO interface which, in turn, causes the increase of the 2DEG density.

Turning to the ferroelectric cases, we consider for the magnitude of both P$_{\textrm{up}}$ and P$_{\textrm{down}}$, a polarization of about 0.12 C/m$^2$. As the in-plane unit cells of the heterostructures are fixed to that of BaSnO$_3$, the latter value is lower than the polarization in pristine BaTiO$_3$ ($\sim$0.29 C/m$^2$~\cite{Fang+long}) but comparable to that of strained bulk BaTiO$_3$ (0.10 C/m$^2$~\cite{Fang+long}). The obtained electron charge densities within the BaSnO$_3$ side are 0.49 e/a$^2$ and 0.26 e/a$^2$ for P$_{\textrm{down}}$ and P$_{\textrm{up}}$, respectively. Comparing with the paraelectric case, we clearly see that the P$_{\textrm{down}}$ case accumulates electronic charge across the LIO/BSO interface and enhances the charge density, while the P$_{\textrm{up}}$ case results in charge depletion and thus a lower 2DEG density.

To get a better understanding of the mechanism behind, we analyze the band structures and the local densities of states (LDOS) of the different systems and summarize our findings in Fig.~\ref{Fig-2}. For the paraelectric case, the resulting two-dimensional hole gas (2DHG) is mainly confined at the BTO/LIO interface with some extension to the BaTiO$_{3}$ layers. When we set the polarization toward the LIO/BSO interface (P$_{\textrm{down}}$  case), the dipole induced within the BaTiO$_3$ block causes an upward shift of the valence-band edge starting from the LaInO$_3$ side (left panel). As such, the hole charge is enhanced and localizes at the BaTiO$_3$ surface rather than at the BTO/LIO interface. This result shows up in an increase of the electronic charge density within the BaSnO$_3$ block ($\sim$0.49 e/a$^2$). Interestingly, comparing the 2DEG distribution with the paraelectric case (Fig.~\ref{Fig-1}), we find that it is more confined and accumulates near the LIO/BSO interface. Switching the polarization direction from the LIO/BSO interface outward (P$_{\textrm{up}}$ case), induces a dipole within the BaTiO$_3$ block, but in contrast to the P$_{\textrm{down}}$ case, it causes a downward shift of the valence-band edge starting from the LaInO$_3$ side of the BTO/LIO interface. Therefore, the hole charge is enhanced and localizes at the BTO/LIO interface, thereby increasing the 2DEG charge density. However, a part of the electron charge transfer is toward the BaTiO$_3$ surface to compensate its ferroelectric dipole. This process drastically reduces the 2DEG density within the BaSnO$_3$ block to about 0.26 e/a$^2$ compared to that of the paraelectric case (0.37 e/a$^2$). Overall, we conclude that the ferroelectric layer with P$_{\textrm{up}}$ polarization allows for depleting the 2DEG charge from the BaSnO$_3$ side.

%%%%%%%%%%
\begin{figure}
 \begin{center}
\includegraphics[width=.49\textwidth]{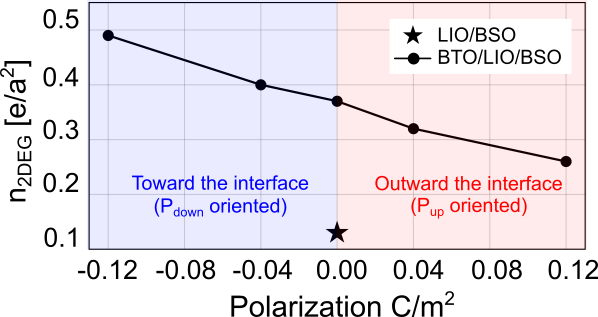}%
\caption{Variation of the 2DEG density in the BTO/LIO/BSO heterostructure within the BaSnO$_3$ block as a function of polarization of the ferroelectric BaTiO$_3$ layer. The star refers to the 2DEG density of the pure LIO/BSO interface.}
\label{Fig-3}
\end{center}
\end{figure}
%%%%%%%%%%%%%%%%%%%%%%%%%%%%%%%%%%%%%%%%%%%%%%%%%

To validate the previous findings, we vary the magnitude of the ferroelectric polarization and explore how it affects the 2DEG density within the BaSnO$_3$ block. As depicted in Fig.~\ref{Fig-3}, the 2DEG charge density increases (decreases) gradually by increasing the magnitude of the ferroelectric polarization, P$_{\textrm{down}}$ (P$_{\textrm{up}}$). This trend indicates that one can accumulate/deplete the 2DEG significantly \cite{note}. Overall, the tunability of the 2DEG discussed here for the BTO/LIO/BSO combination is inline with the experimental prediction for Pb(Zr,Ti)O$_3$/LaAlO$_3$/SrTiO$_3$ \cite{tra+2013AM}.

Let us finally comment on the BaTiO$_3$ thickness which can also be used to tune the induced electric field. Due to the computational complexity, we have limited our calculations to four BaTiO$_3$ unit cells. To understand, nevertheless, how the thickness impacts the 2DEG density, we reduce it to three BaTiO$_3$ unit cells (details will be provided in a forthcoming publication~\cite{Fang+long}). The polarization is chosen such to achieve a similar electric field as in the previous case. We find that the 2DEG charge decreases, \ie it should conversely be enhanced with an increasing number of BaTiO$_3$ layers. All this implies that one can effectively tune the 2DEG density.

In conclusion, based on first-principles calculations, we have predicted a controllable 2DEG at the interface between the polar LaInO$_3$ perovskite and the nonpolar BaSnO$_3$ counterpart by adding a functional ferroelectric BaTiO$_3$ layer. A polarization pointing toward the interface enhances its polar discontinuity, causing an accumulation of the 2DEG at the BaSnO$_3$ side, \ie enhancing its density. Switching the ferroelectric polarization outward of the interface depletes the 2DEG charge. Focusing on the ferroelectric side, we have found that the 2DEG density increases with its thickness. As epitaxial ferroelectric layers used experimentally are usually about tens of nanometers or more~\cite{kim+2013AM,wang+2018ACS}, they are expected to give rise to a substantially higher 2DEG density. In addition, thicker layers allow for reaching higher polarization magnitudes. Moreover, we like to provide and outlook regarding other ferroelectric functional layers like Pb(Zr,Ti)O$_3$ and (K,Na)NbO$_3$. The latter offers the possibility for tuning its lattice parameter by adjusting the rate of K and Na atoms~\cite{saito+2004JAP}, thereby minimizing the lattice mismatch with the polar material. Overall, ferroelectric layers provide an effective way for controlling the characteristics of a 2DEG in terms of density and spatial distribution, which can be generalized to other heterostructures consisting of ferroelectric, polar, and nonpolar components. Such heterostructures can be turned into practical applications, like in ferroelectric FETs.

\section*{Data availability} 
Input and output files can be downloaded free of charge from the NOMAD Repository~\cite{drax-sche19jpm} at the following link: \url{https://dx.doi.org/10.17172/NOMAD/2022.01.11-1}.

%%%%%%%%%%%%%%%%%%%%%%%%%%%%%% 
%%%%%%%%%%%%%%%%%%%%%%%%%%%%%%%%%%%%%%%%%%
\section*{Acknowledgment} 

L.F. and W.R. are grateful for the support from the National Natural Science Foundation of China (51861145315, 11929401, 12074241), the Independent Research and Development Project of State Key Laboratory of Advanced Special Steel, Shanghai Key Laboratory of Advanced Ferrometallurgy, Shanghai University (SKLASS 2020-Z07), the Science and Technology Commission of Shanghai Municipality (19DZ2270200, 19010500500, 20501130600), and the China Scholarship Council (CSC). This work was supported by the project BaStet (Leibniz Senatsausschuss Wettbewerb, No. K74/2017) and was performed in the framework of GraFOx, a Leibniz Science Campus, partially funded by the Leibniz Association. We acknowledge the North-German Supercomputing Alliance (HLRN) for providing HPC resources (project bep00078 and bep00096). W.A. and L.F. thank Martin Albrecht, Martina Zupancic (Leibniz-Institut f\"{u}r Kristallz\"{u}chtung, Berlin), Dmitrii Nabok (Humboldt-Universit\"{a}t zu Berlin), Chen Chen (Shanghai University) and Kookrin Char (Seoul National University) for fruitful discussions. 
%%%%%%%%%%%%%%%%%%%%%%%%%%%%%%%



\begin{thebibliography}{33}
\bibitem [{not()}]{note}%
	\BibitemOpen
	\href@noop {} {}\bibinfo {note} {More points between 0 and 0.12 C/m$^2$
		without relaxation confirm the linear trend. More details will be provided in
		a forthcoming publication~\cite{Fang+long}.}\BibitemShut {Stop}%
\end{thebibliography}
\end{document}